\documentstyle[vsolj02,graphicx,natbib]{article}

\def\xxinput#1{\input#1}

\xxinput{vsolj02.sty}

\RequirePackage[T1]{fontenc}
\RequirePackage{bm}

\def\cite{\citealt}
\setcitestyle{aysep={}}

\begin{document}

\title{A code for two-dimensional frequency analysis using}
\vskip -2mm
\title{the Least Absolute Shrinkage and Selection Operator (Lasso)}
\vskip -2mm
\title{for multidisciplinary use}

\author{Taichi Kato$^1$}
\author{$^1$ Department of Astronomy, Kyoto University,
       Sakyo-ku, Kyoto 606-8502, Japan}
\email{tkato@kusastro.kyoto-u.ac.jp}

\begin{abstract}
In \citet{kat12perlasso}, we introduced the Least Absolute
Shrinkage and Selection Operator (Lasso) method, a kind of
sparse modeling, to study frequency structures of
variable stars.  A very high frequency resolution was
achieved compared to traditional Fourier-type frequency
analysis.  This method has been extended to two-dimensional
frequency analysis to obtain dynamic spectra.  This
two-dimensional Lasso frequency analysis yielded
a wide range of results including separation of
the orbital, superhump and negative superhump signals
in Kepler data of SU UMa stars.  In this paper, I briefly
reviewed the progress and applications of this method.
I present a full R code with examples of its usage.
This code has been confirmed to detect the appearance
of the orbital signal and the variation of the spin
period after the eruption of the nova V1674 Her.
This code also can be used in multidisciplinary purposes
and I provide applications to analysis of avian vocalizations.
I found fine structures in the call of the Eurasian wren
(\textit{Troglodytes troglodytes}), which is likely to
be used species identification.  This code would
be a new tool in studying avian vocalizations
with high temporal and frequency resolutions.
Interpretations of the power spectra of avian vocalizations
will also be helpful in interpreting the power spectra
of variable stars.
\end{abstract}

\section{Introduction}\label{sec:intro}

   Frequency analysis is essential in many fields of science.
In astronomy (I particularly deal with period/frequency analysis
of variable stars), many types methods have been used.
Fourier-type methods include discrete Fourier transform \citep{DeemingDFT},
least-squares fit of sinusoids for unevenly spaced data
(Lomb-Scargle Periodogram, \cite{LombScargle}; \cite{hor86periodanalysis};
\cite{zec09LombScargle}) and CLEAN algorithm \citep{CLEAN}.
They are related to a variety of Cohen's class transformations
for evenly sampled data, which include short-time Fourier transform
(STFT), wavelet transform and
Wigner-Ville distribution \citep{hla08timefrequency}.
These methods give generally the same frequency resolution
given by the Heisenberg-Gabor limit.

   There are also methods by evaluating dispersions
in phase-sorted or phase-binned data.  This class of
approach includes the string-length method \citep{dwo83stringlength},
Lafler-Kinman's method \citep{LaflerKinman},
Phase Dispersion Minimization (PDM, \cite{PDM}) and
Analysis of Variance (AoV, \cite{sch89AoV}).
These methods give an estimate with a smaller error
than Cohen's class transformations
for the frequency/period when a single signal is present
\citep{fer89error,Pdot2}.
These methods are advantageous over Cohen's class transformations
for a non-sinusoidal signal, which is usual for variable stars.

   In recent years, a completely new approach in
Fourier-type analysis has developed with an assumption that
only small number of frequencies are present at
the same time (condition usually called ``sparse'').
This is an application of a rapidly
developing field of compressed sensing
(e.g. \cite{don06compressedsensing}).  Under this
assumption, we can, in principle, break the Heisenberg-Gabor
limit [A simple explanation why the Lasso-type analysis can overcome
the Heisenberg-Gabor limit in a different field of science
can be found in \citet{her09radarCompressedSensing}].
In \citet{kat12perlasso},
we applied the least absolute shrinkage and selection operator
(Lasso), a form of compressed sensing to frequency analysis
of astronomical time-series observations (also called
``sparse Fourier'' analysis).
This method has a resolution superior to Cohen's class
transformations and Lafler-Kinman's class methods,
and is particularly useful in separating
closely spaced multiple frequency components,
such as the orbital signals and superhumps signals
in SU UMa-type dwarf novae.

   For studying frequency variations, two-dimensional
frequency analysis (dynamic spectrum) is used.
For evenly sampled data, STFT is usually used
(e.g. \cite{sti10v344lyr}).
Although \citet{kat12perlasso} presented a code for
single-epoch analysis, two-dimensional frequency analysis
using Lasso has been employed and various results
have been published (\cite{kat13j1924};
\cite{osa13v344lyrv1504cyg};
\cite{kat13j1939v585lyrv516lyr,kat13j1922};
\cite{osa14v1504cygv344lyrpaper3}; \cite{Pdot5};
\cite{ohs14eruma}; \cite{Pdot6}; \cite{pav14ezlyn};
\cite{kat15ccscl}; \cite{Pdot7}; \cite{kat16rzlmi};
\cite{nii21asassn18ey}).  In \citet{osa13v344lyrv1504cyg,
osa14v1504cygv344lyrpaper3}, Lasso analysis played
an important role in demonstrating the conclusion that
the thermal-tidal instability model \citep{osa89suuma}
is the only viable model to explain superoutbursts
in SU UMa stars.
In these series of papers, we have shown that Lasso analysis
is advantageous in several respects: (1) the resultant signal
is very sharp, (2) the frequency resolution is very high,
and (3) the result is less affected by uneven sampling
than in Fourier analysis.  The last point is particularly
important in real astronomical data, in which observations
can usually be obtained unevenly in time.
Comparisons of spectrograms between STFT and Lasso-based
Fourier analysis are shown in \citet{osa13v344lyrv1504cyg,
kat13j1939v585lyrv516lyr,kat13j1924}.
The advantage of Lasso or sparse modeling in Fourier-type
application has been extended to super-resolution imaging with
radio interferometry by \citet{hon14superresolution}.

   Considering that two-dimensional frequency analysis using
Lasso is useful not only in astronomy but also in other
fields of science, I present a multi-purpose full code
to calculate and draw two-dimensional Lasso spectral analysis.
The code was originally written for \citet{kat13j1924},
and has been modified here for more general purposes.

\section{Methods}

   The simplified mathematical basis of our Lasso analysis
is given here.  For a more complete treatment
and comparison with other frequency determination
algorithms, please refer to \citet{kat12perlasso}.
This mathematical basis is not always necessary to use
the code, and readers can skip this section to understand
actual applications and to learn how to use the code.

   The mathematical formulation,
which is based on \citet{tan10compressedsensing}.
The data (in amplitude) have $Y(t_i)$ with
times of observations at $t_i$.  The mean of $Y$ is
set to be zero. 
The observation can be expressed as a sum of signal ($Y_s$) and
random errors ($n$) :
  \begin{equation}\label{equ:obserror}
  Y_i = Y(t_i) = Y_s(t_i) + n(t_i).
  \end{equation}
The signal is assumed to be composed of a sum of strictly periodic
functions.
Using sine and cosine Fourier components, $Y_s$ can be expressed as :
  \begin{equation}\label{equ:fouriersum}
  Y_s(t_i) = \sum_j{a_j \cos (\omega_j t_i)} + \sum_j{b_j \sin (\omega_j t_i)},
  \end{equation}
where $\omega$ are frequencies and $a$ and $b$ are amplitudes.
This equation can be rewritten as
  \begin{equation}
  \bm{y} = (Y_1, \cdots, Y_N)^{\mathrm{T}}
  \end{equation}
and
  \begin{equation}
  \bm{x} = (a_1, \cdots, a_M, b_1, \cdots, b_M)^{\mathrm{T}},
  \end{equation}
where $N$ and $M$ are number of observations and number of
different $\omega$, respectively.
A set of equations \ref{equ:obserror} and
\ref{equ:fouriersum} can be rewritten as a form of
  \begin{equation}
  \bm{y} = A\bm{x} + \bm{n}
  \end{equation}
using a $2M \times N$ observation matrix $A$ composed of
  \begin{equation}\label{equ:fouriercomp}
  A_{i,j} = \left\{
    \begin{array}{ll}
      \cos (\omega_i t_j), & \mbox{($i \le M$)} \\
      \sin (\omega_{i-M} t_j), & \mbox{($i > M$)}.
    \end{array}
    \right.
  \end{equation}
The vector $\bm{x}_0$ is what to be estimated.

   In Lasso, ``1-norm'' is used, 
which is $||\bm{x}||_1 \equiv \sum_{i=1}^N|x_i|$,
and $\hat{\bm{x}}^{\mathrm{LAR}}$:
  \begin{equation}\label{equ:lassodef}
  \hat{\bm{x}}^{\mathrm{LAR}} = \arg \min_{\bm{x}} (\frac{1}{2N} ||\bm{y} - A\bm
{x}||^2 + \lambda ||\bm{x}||_1),
  \end{equation}
where $||\bm{y} - A\bm{x}||^2$ is the ordinary 
squared sum of residuals, is chosen, and 
$\lambda ||\bm{x}||_1$ is the $\ell_1$-norm penalty function
with a parameter $\lambda$ having a value of $\lambda \ge 0$.
$\arg \min_{\bm{x}} f(\bm{x})$ means the value of $\bm{x}$
which minimizes $f(\bm{x})$.
This estimate becomes identical with a least-squares estimation
at $\lambda = 0$.  It has been known that this $\ell_1$
regularization provides a sparse solution (small number of
non-zero elements in $\bm{x}_0$) in most cases
(cf. \cite{don06compressedsensing,can06signalrecovery})
and there is a fast algorithm, known as least angle regression
(LAR) by \citet{LARS} to solve this problem.

   In this Lasso analysis, I used LAR implementation
{\bf lars} package by \citet{LARS} on R software\footnote{
   The R Foundation for Statistical Computing:
   $<$http://cran.r-project.org/$>$.
} combined with {\bf glmnet}
(generalized linear model via penalized maximum likelihood;
\cite{glmnet}) as a wrapper.

\section{Application to frequency analysis of variable stars}\label{sec:varstars}

\subsection{Usage of the code}

\begin{figure*}
\begin{center}
\includegraphics[angle=-90,width=11cm]{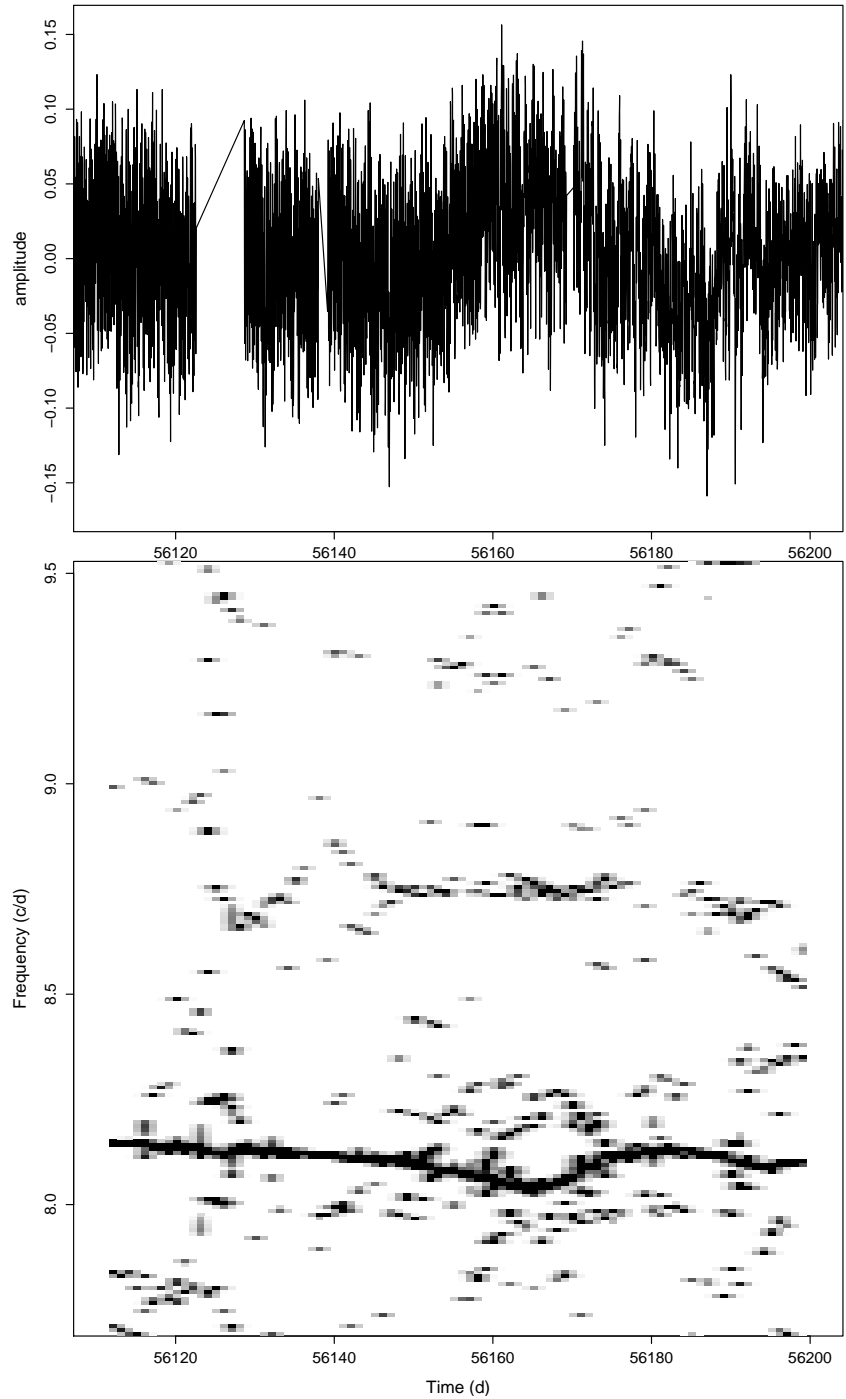}
\caption{Example of graphical representation of
  two-dimensional Lasso analysis on R using the code presented
  in this paper.  Kepler data of KIC 8751494 were used
  (see \cite{kat13j1924}).
  The signal at 8.74~c/d is the orbital signal.
  The varying signal between 8.0 and 8.2~c/d is
  the superhump signal.  See text how to produce this figure.
}
\label{fig:kicout}
\end{center}
\end{figure*}

   Since previous applications in research of variable stars
are already given in section \ref{sec:intro},
I only provide an example to run a code.
Assuming that readers have the data of KIC 8751494
\citep{kat13j1924} as a form of a plain text with
BJD$-$2400000 and magnitudes ready to read with R,
the analysis can be done with:
\begin{verbatim}
source("lassobase.R")
dat <- read.table("datafile")

x <- dat$V1 - mean(dat$V1)
le <- lowess(x,dat$V2,f=0.05)
res <- dat$V2 - le$y
tab <- data.frame(V1=dat$V1,V2=res-mean(res))

pgm <- getpergrmlasso2(tab,56106,56205,1/0.130,1/0.105,200,10,1)
\end{verbatim}
The second part is for prewhitening the data using
the locally-weighted polynomial regression (LOWESS: \cite{LOWESS}).
This part needs to be modified depending on the complexity
of the light curve.  This is particularly important
for analyzing light curves of dwarf novae.
In papers such as \citet{osa13v344lyrv1504cyg}, we used
a very small smoother span (f in function \verb|lowess|)
and a small delta in the LOWESS regression to remove
the global patterns of outbursts in dwarf novae and
processed the entire data as a whole.
This prescription led to favorable results
for many dwarf novae despite that the amplitudes of
the target signals (superhumps, orbital variations and others)
are greatly reduced.  To conserve the amplitudes of
the target signals, it is better to divide the entire
light curve to shorter segments so that the LOWESS regression
requires a larger f (typically larger than 0.1) to obtain
sufficiently prewhitened segments but with well-conserved
target signals.  These short segments are combined to
obtain the prewhitened entire data for the two-dimensional
Lasso analysis.

   The second and third arguments for \verb|getpergrmlasso2|
gives the time range, and the fourth and fifth for
the frequency range (c/d in this case).  The sixth argument
200 is the number of frequency bins, the seventh 10 is
the window length and the final 1 is the shift value for
the windows.
The resultant \verb|pgm| can be saved using \verb|save|
function in R for further analysis or graphical representation.
The result \verb|pgm| can be readily seen on graphical window by
\begin{verbatim}
drawpgmlasso(pgm,-2.7,-13,-6,3)
\end{verbatim}
The second argument $-$2.7 is $\log_{10} \lambda$ and
the third and fourth $-$13 and $-$6 are
the display range of the power strength (in logarithmic scale)
and the final 3 is time bins to smear
(use 0 when no smearing is needed).
The R graphical window would look like figure \ref{fig:kicout}.
The $\lambda$ value can be optimized by looking at the resultant
two-dimensional spectrogram by changing the gray scale
(by changing the display range of power strength),
examining whether target frequencies are well visible
above the background noise.
The list of different values of $\lambda$ by \textbf{lars} is
stored in the vector \verb|pgm$lst[[i]]$lassopow$lambda|, where
\verb|i| is the bin number (1 $\leq$ \verb|i| $\leq$
\verb|length(pgm$lst)|), and
one can find an adequate range of $\log_{10} \lambda$ by using
these values.
One can also draw a spectrogram of a specified element
(\verb|n|) within a series of different $\lambda$
given by \textbf{lars} by setting the argument \verb|loglambda|
to \verb|FALSE|:
\begin{verbatim}
drawpgmlasso(pgm,n,-13,-6,3,loglambda=FALSE)
\end{verbatim}
(\verb|n| must be an integer and 1 $\leq$ \verb|n| $\leq$
\verb|length(pgm$lst[[i]]$lassopow$lambda)|).
By changing \verb|n|, one can easily find the optimal
$\lambda$.

   Although one can use the popular cross-validation technique
to obtain $\lambda$ expressing the most regularized model
(see \cite{kat12perlasso}), this is not always optimal
for actual applications.  This is probably because the model
using sine and cosine functions is too simple for
describing the real variation.

\begin{figure*}
  \begin{center}
  \includegraphics[angle=0,width=10cm]{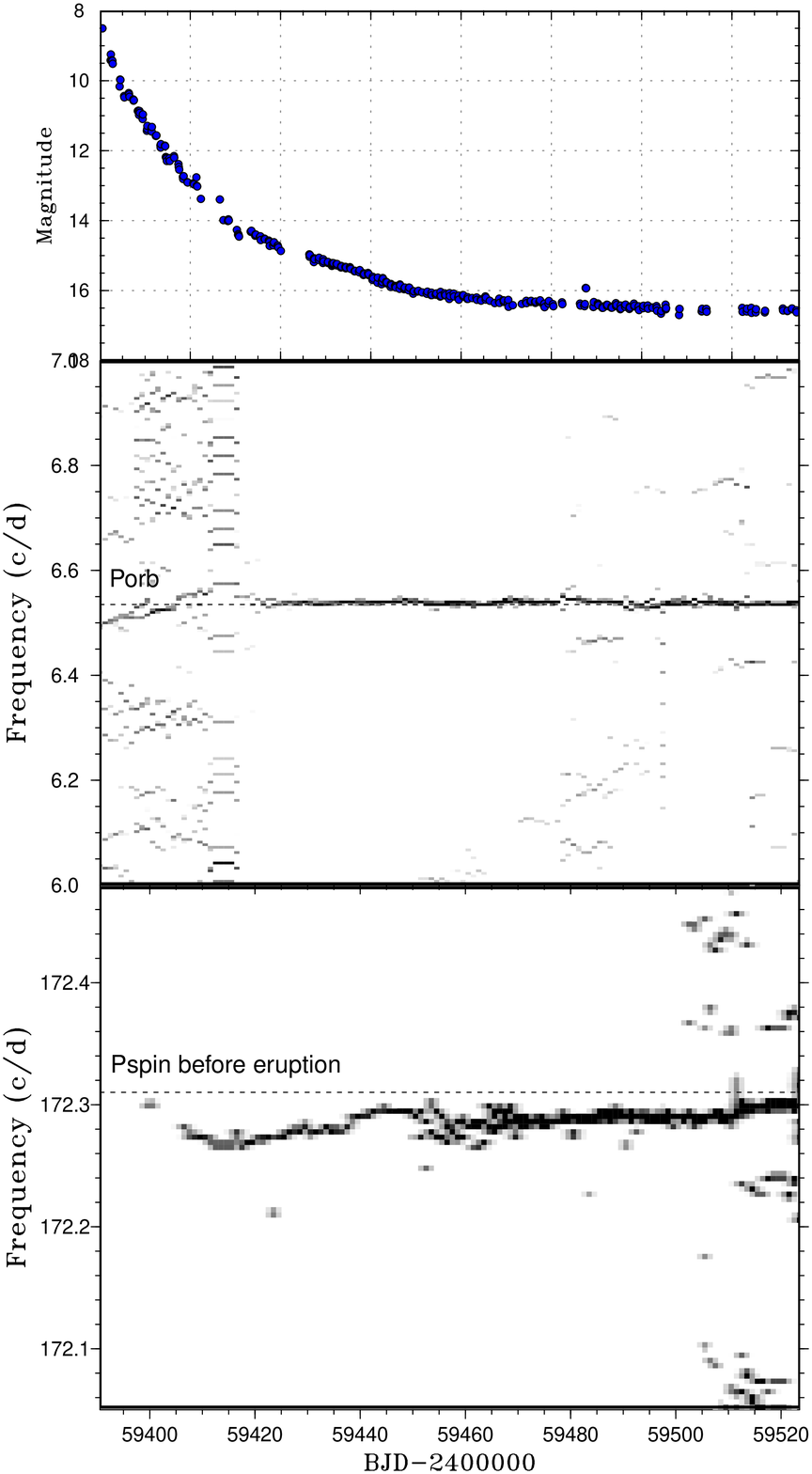}
  \caption{Lasso two-dimensional power spectrum
  of the nova V1674 Her.  The data were from observations
  reported to VSNET Collaboration \citep{VSNET}.
  (Upper) Light curve.  The data were binned to 0.1~d.
  (Middle) Lasso power spectrum around the orbital period
  The orbital signal (6.54~c/d, corresponding to a period
  of 0.1530~d) has been present since BJD 2459420.
  (Lower) Lasso power spectrum around the spin period.
  The window length and shift value for computing
  these spectra are 20~d and 1~d, respectively.
  }
  \label{fig:v1674herspec}
  \end{center}
\end{figure*}

\subsection{Application to the super-fast nova V1674 Her}

   I give another example of a Lasso two-dimensional power spectrum
of the fastest known nova V1674 Her.  This nova was
discovered by Seiji Ueda on 2021 June 12,\footnote{
  $<$http://www.cbat.eps.harvard.edu/unconf/followups/J18573095+1653396.html$>$.
} and is the fastest ever recorded classical nova
with $t_2\sim 1.2$~d \citep{qui21v1674her}.
This nova was reported to show coherent pulses with
a period of 501.428~s in Zwicky Transient Facility (ZTF; \cite{ZTF})
data before the eruption \citep{mro21v1674heratel14720} and
this period is considered to reflect the spin period
of white dwarf.  In addition to this period, orbital modulations
with a period of 0.15302(2)~d was identified after
the nova eruption \citep{shu21v1674heratel14835,
pat21v1674heraatel14856,shu21v1674her}.
\citet{pat21v1674heraatel14856} reported a spin period
of 501.52(2)~s after the eruption.  This value is
0.09(2)~s longer than the value before the eruption
and is consider to be a consequence of the nova eruption,
either by an increase in the moment of inertia
or by a loss of mass that carried away angular momentum
\citep{dra21v1674her}.

   I examined whether these signals can be seen
in Lasso two-dimensional power spectra.  I used the data
reported to VSNET Collaboration \citep{VSNET}
(observers: Tonny Vanmunster, Michael Richmond,
Stephen Brincat, Charles Galdies and Franz-Josef Hambsch).
The analysis presented here is preliminary and is
planned to appear in VSNET Collaboration in prep.
with more details.
In the middle panel of figure \ref{fig:v1674herspec},
the signal of the orbital period appeared around
BJD 2459420 ($\sim$40~d after the eruption).
There was some hint of modulations with a frequency
slightly lower than that of the orbital signal
before BJD 2459410.  This signal has a period longer
than the orbital period and may be some sort of
superhumps, although the data were not sufficient
to draw a conclusion.
In the lower panel of figure \ref{fig:v1674herspec},
a spin frequency was detected below the one observed before
the eruption.  This has confirmed the results
by \citet{dra21v1674her}.  Although period variations
of the order of 0.1--0.3~s suggested from X-ray observations
\citep{dra21v1674her} were not apparent, the Lasso
analysis suggests that the spin frequency had a tendency to
return to the frequency before the nova eruption.
This may be supportive of an idea that the decrease in
the spin frequency was the result of the increase
in the moment of inertia due to the nova explosion.

\section{Application to frequency analysis of vocalization of birds}

\subsection{General introduction and usage of the code}

   In analyzing vocalizations of birds, two-dimensional
Fourier power spectra (sonograms) have been widely used
(e.g. \cite{mar04birdsong}).  STFT is usually used to construct
two-dimensional spectra.
Both open-source, cross-platform audio software
\textbf{audacity}\footnote{
  $<$https://www.audacityteam.org/$>$
}, or software specially designed for analysis of
vocalizations of birds \textbf{Raven Lite}\footnote{
  $<$https://ravensoundsoftware.com/software/raven-lite/$>$.
} can produce research-grade sonograms.
On R, \textbf{dynspec} function in \textbf{seewave}
package\footnote{
  $<$https://cran.r-project.org/web/packages/seewave/index.html$>$.
} can produce a two-dimensional STFT power spectrum
both as a form of matrix and graphical representation.
They are sufficient to study sonograms of vocalizations of
birds in many situations.

   Although this method is well-established
in bioacoustic studies of birds, time resolutions or
frequency resolutions limited by the Heisenberg-Gabor limit
(section \ref{sec:intro}; see also \cite{bee88uncertainty})
could become a problem especially when the durations
of sounds are short.  In analysis of song features,
frequency tracking or methods on focusing on
the main frequency have been
applied to overcome this difficulty
\citep{gal12senderreceiver,sto12chirp,sto14chirp}.
Such methods, however, can have difficulties in handling
sounds with multiple frequency components.
Furthermore, it has been demonstrated that ultrafast
(over 200~Hz) movements of vocal muscles control birdsong
in certain species \citep{ele08superfastmuscles}.
In such vocalizations, the duration of each note is
about 1~msec.  Even the optimal selection of parameters
can give a frequency resolution of 1~kHz by the traditional
STFT method, and it may not be adequate for tracing
the variation of frequencies within individual notes.
\citet{ele08superfastmuscles} indeed employed the sound
strength as a tracer of vocalization, but if we could
trace the frequency variation, it would provide more
information about the characteristic of notes and
mechanisms to produce and control these sounds.

   In order to analyze vocalizations of birds by
the present code, the sound data either in
linear pulse code modulated audio (LPCM) format
or conventional compressed format such as
MP3 need to be read just like time-series data for variable stars.
For reading LPCM wav format files, 
\textbf{readWave} function in \textbf{tuneR} package\footnote{
  $<$https://cran.r-project.org/web/packages/tuneR/index.html$>$.
}
is useful.  The actual code would be like:
\begin{verbatim}
library(tuneR)

source("lassobase.R")

s <- readWave("wren.wav",from=0,to=2,units="seconds")
y <- s@left
rate <- s@samp.rate
y <- y - mean(y)
x <- (1:length(y))/rate * 1000
d <- data.frame(V1=x,V2=y)

pgm <- getpergrmlasso2(d,1390,1425,2,16,100,2,0.4)
\end{verbatim}
In this example, I used a call of an Eurasian Wren
(\textit{Troglodytes troglodytes}) recorded locally in Japan
using a SONY recorder ICD-SX1000 in the LPCM mode (96 kHz/24-bit).
The usage of \verb|getpergrmlasso2| would be self-evident
when compared to section \ref{sec:varstars}.
In sound analysis, the times are given in msec and
frequencies are given in kHz.
In order to obtain a graphical representation
\verb|drawpgmlasso| is used as in analysis of
variable stars.  In sound analysis, the amplitudes are
usually much larger than in variable stars (of course,
depending on the definition of the amplitudes) and
$\lambda$ and the display range have larger values.
In this case,
\begin{verbatim}
drawpgmlasso(pgm,2.7,7,12,3,xlab="Time (msec)",ylab="Frequency (kHz)")
\end{verbatim}
yielded a sonogram of optimal quality.
The parameters \verb|xlab| and \verb|ylab| are necessary
to override the default values designed for study of
variable stars.

   For MP3 files, \textbf{readMP3} in \textbf{tuneR} package
can be used instead:
\begin{verbatim}
library(tuneR)

source("lassobase.R")

s <- readMP3("sound.mp3")
y <- s@left
rate <- s@samp.rate
y <- y - mean(y)
x <- (1:length(y))/rate * 1000
d <- data.frame(V1=x,V2=y)
\end{verbatim}
and the usage of \verb|drawpgmlasso| is the same as the case
as with LPCM recordings.  I must note, however, the quality
of MP3 files are highly variable depending on recording
devices and the MP3 compression quality.  Although the use of
LPCM recordings is highly recommended, MP3 files recorded
by suitable recorders with high-quality compressions
(such as at 320~kbps) can be used in frequency analysis
usually without a problem.

   In the following subsections, I describe selected results
of analysis of vocalizations of birds by this code.
They may be useful not only to biologists but also
to astronomers to interpret the features in Lasso spectrograms.
The data used for this analysis were either recorded
by myself in the LPCM mode or from xeno-canto public archive\footnote{
  $<$http://www.xeno-canto.org/$>$.
} (recording numbers are given after XC).
The current archive of xeno-canto consists of MP3
recordings and I used relatively high-quality ones.

\begin{figure*}
  \begin{center}
  \includegraphics[angle=0,width=10cm]{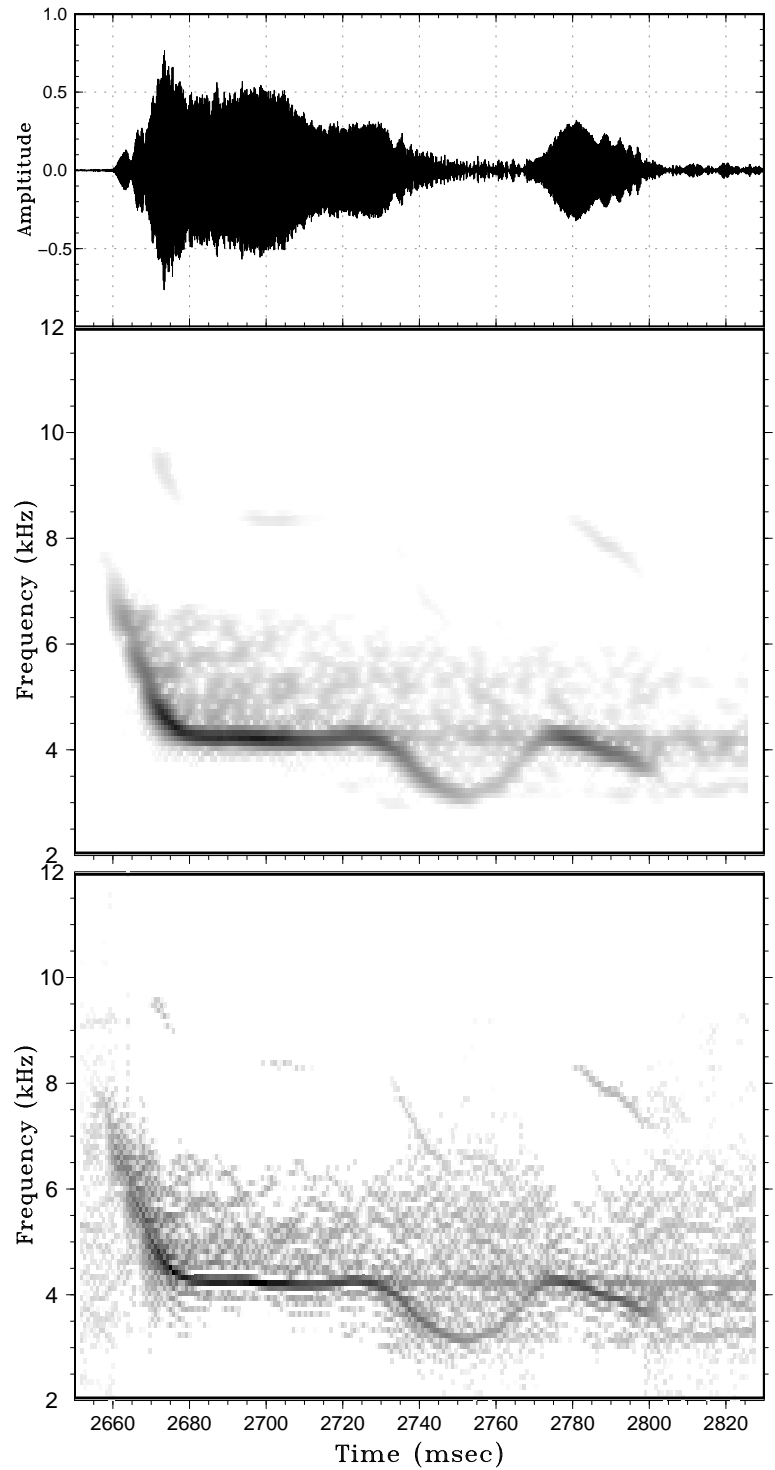}
  \caption{Comparison of Lasso and STFT
  two-dimensional power spectra of a Common Chiffchaff
  \textit{Phylloscopus collybita} (XC214244 in xeno-canto).
  The time is in msec from the start of the audio file.
  (Upper) Amplitude of the signal.
  (Middle) STFT two-dimensional power spectrum.
  100 frequency bins and an 8~msec window size shifted
  by 0.8~msec were used.  The window size was optimized
  to obtain the best resolution.
  (Lower) Lasso two-dimensional power spectrum.
  100 frequency bins and a 4~msec window size shifted
  by 0.8~msec were used.
  The signal is resolved more sharply than in the STFT analysis.}
  \label{fig:chiffoulassocomp}
  \end{center}
\end{figure*}

\subsection{Birdsong with frequency modulations}

   I used a song of a Common Chiffchaff (\textit{Phylloscopus collybita})
(XC214244) for a comparison between STFT and
Lasso spectrograms (figure \ref{fig:chiffoulassocomp}).
This song contains only one fundamental tone and shows
strong frequency modulations.  Since birdsong usually
has a broader range of frequency modulations than in
many astronomical signals, the advantage of Lasso analysis
in high frequency resolution is less apparent than in
a similar comparison in \citet{kat13j1924}.
The frequencies, however, were detected as much sharper
signals than in the STFT analysis.  This result is as sharp
as those by frequency extraction methods reported in
\citet{sto14chirp}.
Although the second harmonic (first overtone)
was detected in both analyses, the signal was more sharply
detected in the Lasso analysis.  The strength of
the second harmonic tends to increase when the frequency of
the fundamental tone decreases.  This feature is less
apparent in the STFT analysis.
The comparison suggests that both STFT and Lasso methods
can be equally used in analysis of song with relatively
slow frequency modulation.  The difference may be more
striking when more rapid frequency modulations are involved.

   This result for the Lasso analysis can be reproduced
by the following code:
\begin{verbatim}
library(tuneR)

source("lassobase.R")

s <- readMP3("XC214244.mp3")
y <- s@left
rate <- s@samp.rate
y <- y - mean(y)
x <- (1:length(y))/rate * 1000
d <- data.frame(V1=x,V2=y)

pgm <- getpergrmlasso2(d,2650,2830,2,12,100,4,0.8)

drawpgmlasso(pgm,0,0,8,5,xlab="Time (msec)",ylab="Frequency (kHz)")
\end{verbatim}

\begin{figure*}
  \begin{center}
  \includegraphics[angle=0,width=11cm]{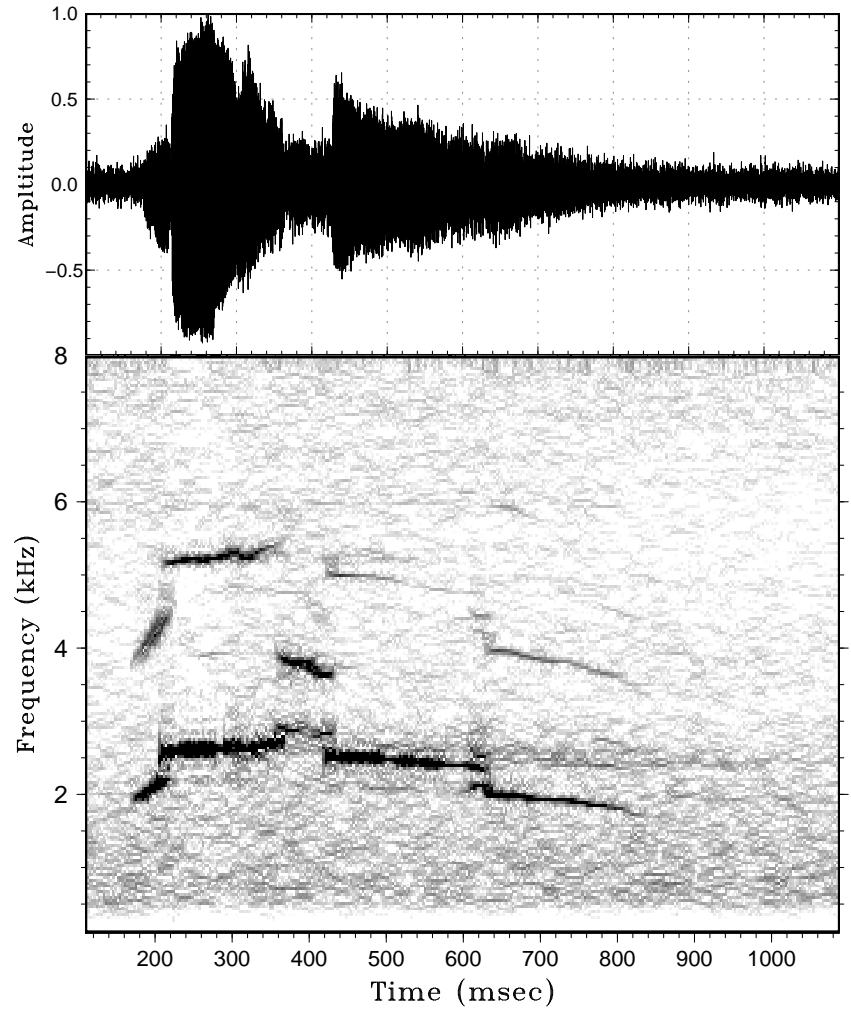}
  \caption{Lasso two-dimensional power spectrum
  of a call of a Crested Honey Buzzard
  (\textit{Pernis ptilorhynchus}).
  (Upper) Amplitude of the signal.
  (Lower) Lasso two-dimensional power spectrum.
  200 frequency bins and a 20~msec window size shifted
  by 3~msec were used.
  The complex structure of the vocalization is well depicted
  together with the prolonged component around 2.5~kHz.
  }
  \label{fig:hachispec}
  \end{center}
\end{figure*}

\subsection{Analysis of a call of a Crested Honey Buzzard}

   In this case, I deal with a call with structures.
The Crested Honey Buzzard (\textit{Pernis ptilorhynchus})
is a diurnal raptor species.  The call is described as ``high,
four-note whistle \textit{wee hey wee hey} or
\textit{pii-yoo pii-ee}, or a whistled scream
\textit{kleeeur} during breeding season''
\citep{BirdsofEastAsia}.\footnote{
  One might see how difficult it is to describe vocalizations
  of a bird by human language!
}
\citet{YachouTaikan} described: ``The voice is around 3~kHz.
There is a short note \textit{pyu} initially, then
a prolonged \textit{iiii-} voice followed by a slightly
lower voice.  The entire voice has a falling intonation.''
(translated from Japanese by myself).

   Since the sonogram in \citet{YachouTaikan} indicated
the presence of a prolonged voice with a constant frequency
lasting 1~s overlapped on the rest of the notes,
I examined how a Lasso two-dimensional power spectrum
can resolve these features.  The voice was recorded by myself
as in the case of the Eurasian Wren.  The result is shown in
figure \ref{fig:hachispec}.  The prolonged voice at 2.5~kHz
is also seen in this spectrogram up to 1000~msec.
Since this feature is seen in many other recordings of
this species, it would be interesting to consider
the possibilities: (1) the bird continuously
issued this frequency while issuing other frequencies,
or (2) it is an echoed signal of the initial note or
(3) the vibration somehow persisted in the body of the bird.
Although echoing is a promising explanation, this recording
was obtained in an open space and it seemed to me difficult
to consider a structure causing echoes within 150~m
(the distance required from the duration of the signal).
If the possibility (1) is the case, this phenomenon is called
polyphony and it suggests that the bird controls
a pair of muscles of the syrinx (vocal organ of birds)
differently.  Such a case is rarely documented 
in non-passerine birds (songbirds) and would
be worth studying in more detail.

   The entire sonogram is characterized by the variable
fundamental frequency and the second (sometimes the third)
harmonic.  Between 370 and 430 msec, the voice strength became
smaller but with a higher (around 3.8~kHz) frequency
with multiple frequency components.  The entire sequence
appears to match the description of a four-note whistle
by \citet{BirdsofEastAsia}, although many other notes of
the same species are not as complex as this one.

\begin{figure*}
  \begin{center}
  \includegraphics[angle=0,width=10.5cm]{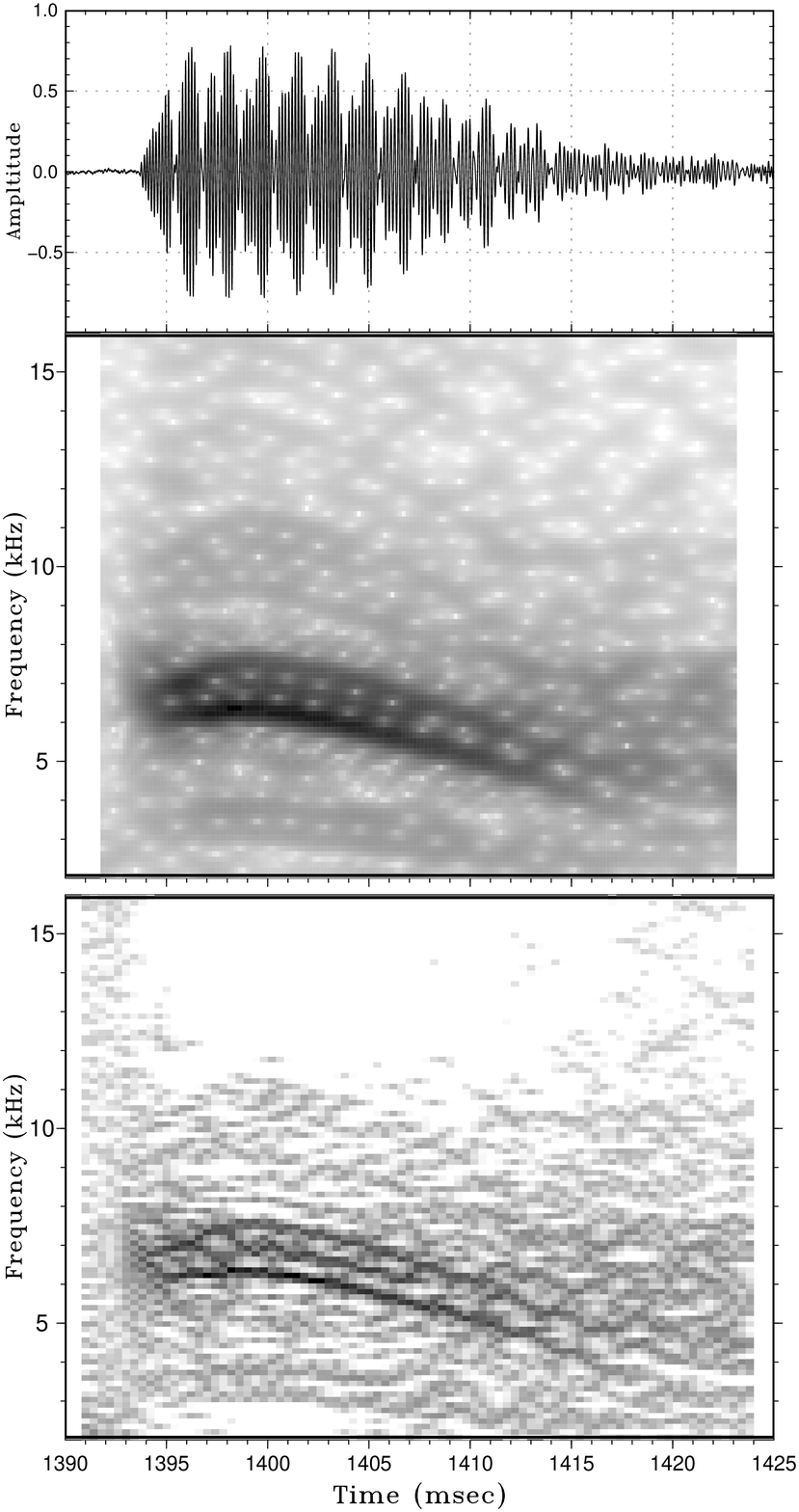}
  \caption{Comparison of Lasso and STFT
  two-dimensional power spectrum of an Eurasian Wren.
  The time is in msec from the start of the audio file.
  (Upper) Amplitude of the signal.
  (Middle) STFT two-dimensional power spectrum.
  100 frequency bins and a 4~msec window size shifted
  by 0.4~msec were used.  The window size was optimized
  to obtain the best resolution.
  (Lower) Lasso two-dimensional power spectrum.
  100 frequency bins and a 2~msec window size shifted
  by 0.4~msec were used.
  Around the strongest part of the sound, the signal is split
  into three frequencies in Lasso analysis.}
  \label{fig:misofoulassocomp}
  \end{center}
\end{figure*}

\subsection{Analysis of call of an Eurasian Wren}\label{sec:wren}

   In this case, I deal with a short call with complex
structures.  A comparison of Lasso and STFT spectra of
a call of an Eurasian Wren is shown in 
figure \ref{fig:misofoulassocomp}.
This call is typical for this species, and I call this
type of call ``usual call''.
This call is short and rich in rapidly varying signals.
Although the STFT analysis, which has been optimized for
resolutions, showed the signal as a barely resolved
band with rapid frequency modulations, the Lasso analysis
clearly showed that this signal is composed of rapidly
varying closely separated frequencies.  
These frequencies varied in parallel.

   Such a pattern of frequencies can be interpreted in
two ways:
(a) They are adjacent higher harmonics.
(b) They are the result of amplitude variation of a single
frequency.  Consider the following equation
in which $f$ and $f_m$ are the original frequency
and the frequency of amplitude modulation, respectively
(usually $f_m \ll f$).  The relative
amplitude of the modulation is given as $a$.
\begin{equation}
{\mathrm signal} = 
\{1 + a \sin (2 \pi f_m t)\} \sin (2 \pi f t)
= \sin (2 \pi f t)
+ \frac{a}{2}
[\cos \{2 \pi (f - f_m) t\} - \cos \{2 \pi (f + f_m) t\}]
\end{equation}
In the resultant power spectrum, two close frequencies
$f \pm f_m$ appear in addition to the central frequency $f$.

   I consider the interpretation (b) more likely
since the amplitudes showed periodic modulations
directly visible to the eyes
(upper panel of figure \ref{fig:misofoulassocomp}),
and both the central and pararell frequencies are present
and the strengths of the pararell frequencies are
almost the same.  By numerical experiments with
artificially modulated data, a 90\% modulation
in the amplitude is required to reproduce the pattern
in the spectrogram of this recording.
This is in agreement with the actual amplitude variation.

   In variable stars, two closely separated
frequencies can cause a beat phenomenon.
In such cases, however, the strengths of the signals are
different and the resultant amplitude modulations due to
the beat phenomenon are smaller than in the present case.
Frequency analysis of variable stars with amplitude
modulations requires attention in interpreting
the spectrogram as in the present case.

\subsection{Different types of calls in the Eurasian Wren}

\begin{figure*}
  \begin{center}
  \includegraphics[angle=0,width=13cm]{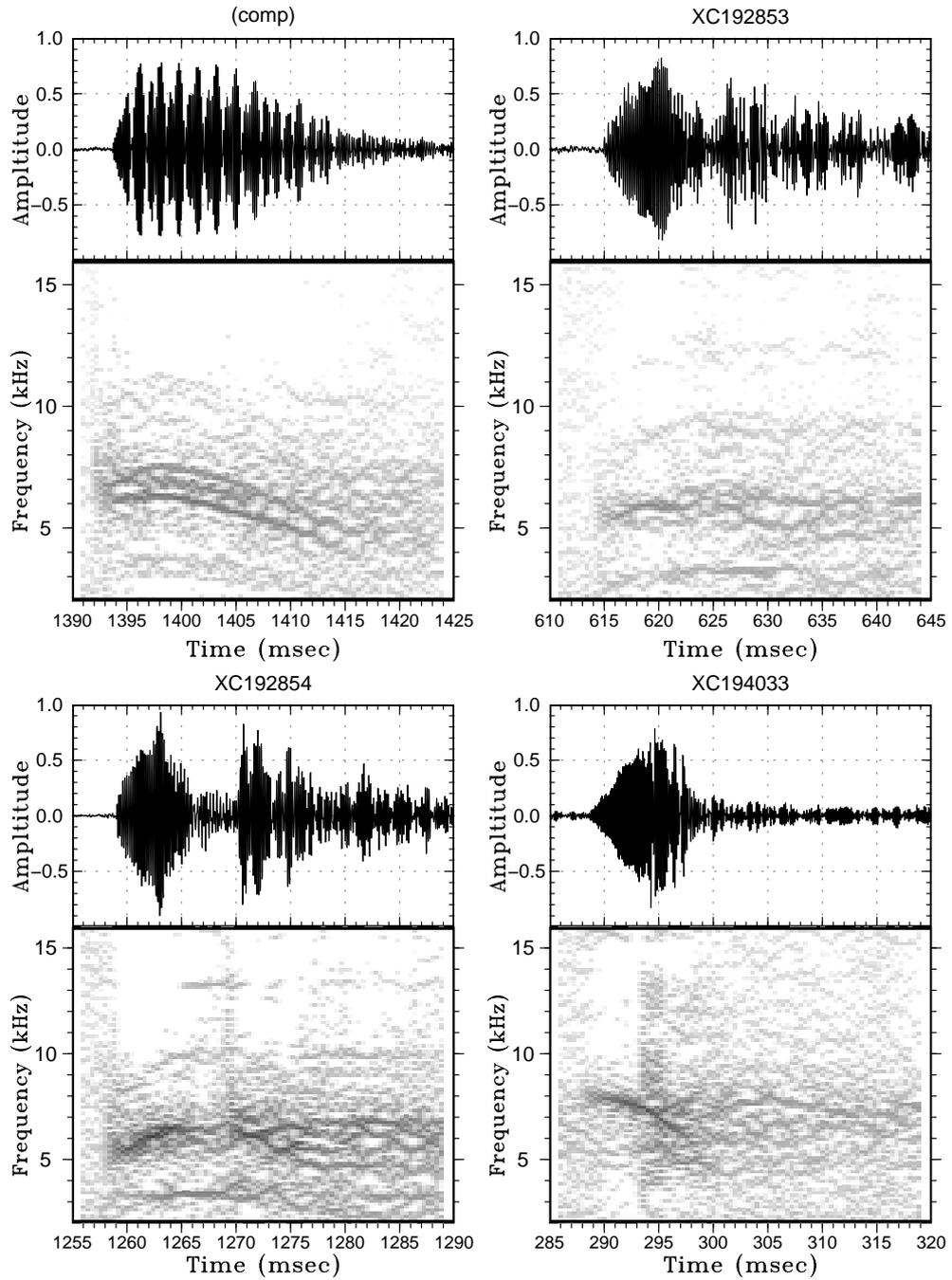}
  \caption{Lasso spectral analysis of different types of calls
  of the Eurasian Wren in Japan.
  The initial one (comp) is the same as the one we used
  in subsection \ref{sec:wren}.}
  \label{fig:misojpncomp}
  \end{center}
\end{figure*}

   Since many avian species have rich vocabularies in calls,
the characteristics of a call of the Eurasian Wren
described in subsection \ref{sec:wren}
may not be applicable to different types of
calls of the same species.
In figure \ref{fig:misojpncomp},
I show two samples (XC192853, XC192854) of the alarm call
and one sample (XC194033) of the trill-type call
(call composed of quickly repeated
short notes) of the same species in Japan.
The same duration (35 msec) is used for analysis.
In the alarm call, the initial part of the note is
characterized by a single rising frequency in contrast to
the usual call.
In the trill-type call, the structure is simple without
closely separated multiple frequencies.  The frequency
of the trill-type call more quickly decreases within
a single note than in the usual call.  We can see that
closely separated multiple frequencies (probably reflecting
amplitude modulation) are not always seen in this species.

\subsection{Comparison between subspecies of the Eurasian Wren}

   I have made a comparative analysis of subspecies
using xeno-canto database to see whether the spectral
characteristics I found (subsection \ref{sec:wren})
are unique to the Japanese subspecies \textit{fumigatus}
or are useful in broadly identifying the species.
Out of 22 subspecies listed in xeno-canto archive,
I analyzed \textit{indigenus}, \textit{zetlandicus}, 
\textit{szetschuanus} and \textit{taivanus}.

\begin{figure}
  \centering
  \includegraphics[angle=0,width=13cm]{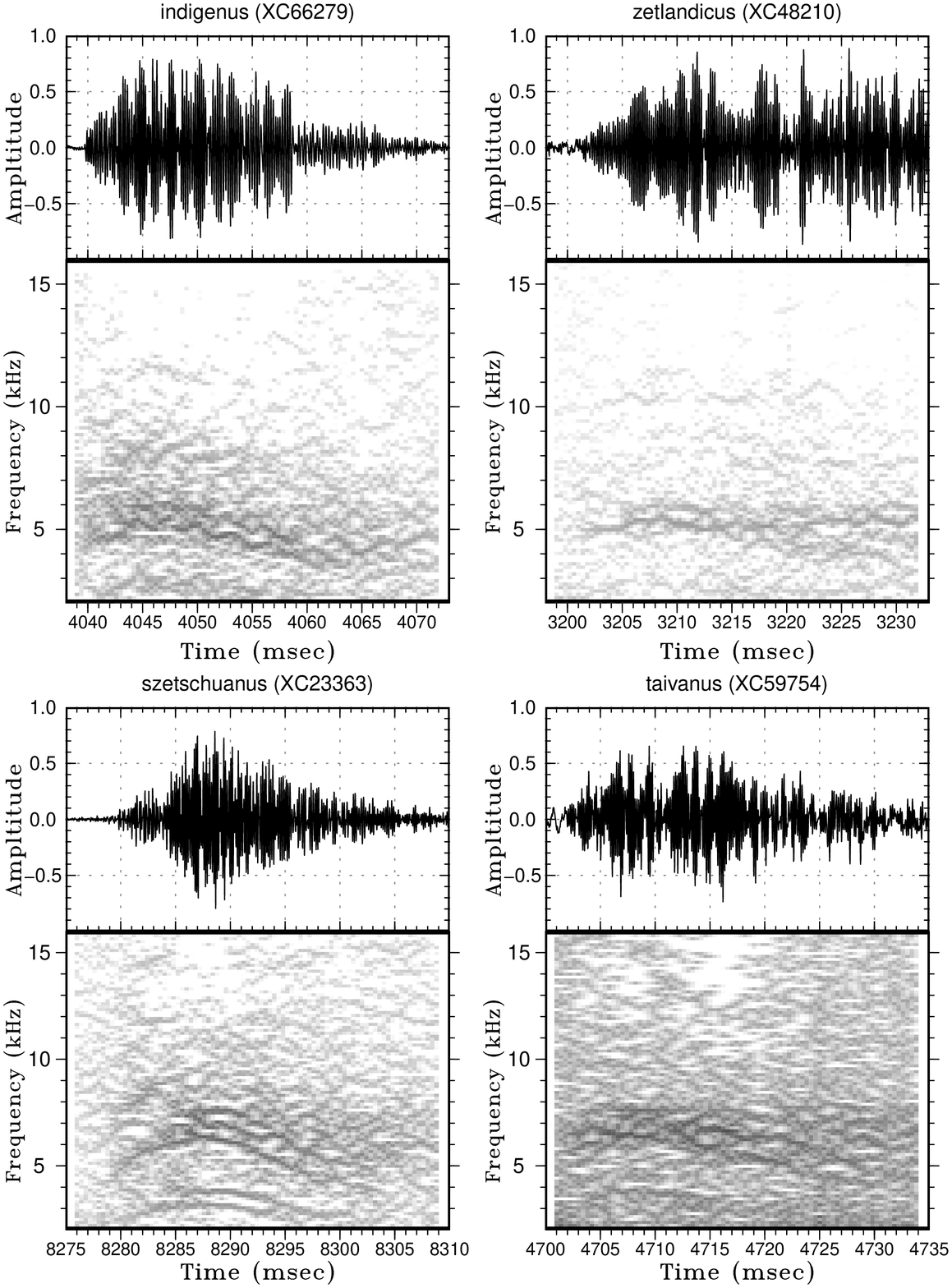}
  \caption{Lasso spectral analysis of different subspecies of
  the Eurasian Wren.}
  \label{fig:misosubcomp}
\end{figure}

   The results are shown in figure \ref{fig:misosubcomp}.
Although one of them (XC23363) is labeled as the alarm call,
all the sounds have characteristics similar to
the usual call recorded in Japan in that they show
closely separated multiple components with decreasing
frequencies.  Since these subspecies are widespread
around the Palearctic, it is highly likely this
frequency structure is common to
other subspecies of the Eurasian Wren, and can be used
for species identification.
The result of XC23363 (\textit{szetschuanus})
even more clearly illustrates the presence of equally spaced
multiple frequencies than in Japanese recordings.
The overall analysis suggests
that the frequency structures of calls in
the Eurasian Wren are highly conserved
between different subspecies.

\section{Code (lassobase.R)}

   This code is also available at
$<$http://www.kusastro.kyoto-u.ac.jp/$\sim$tkato/pub/lassobase.R$>$.

\begin{verbatim}
# The first part is the same as in Kato and Uemura (2012)
# Publications of the Astronomical Society of Japan, 64, 122

library(lars)
library(glmnet)

seqfreq <- function(a,b,...) {
    return(1/seq(1/b,1/a,...))
}

makematlasso <- function(d,p,ndiv) {
    nd <- length(p)
    m <- matrix(0,nrow(d),nd*ndiv)
    for (i in 1:nd) {
        ph <- ((d$V1/p[i]) %% 1)*pi*2
        for (j in 0:(ndiv-1)) {
            m[,i+nd*j] <- sin(ph+pi*j/ndiv)
        }
    }
    return(m)
}

perlasso <- function(d,p,ndiv=2,alpha=1,
            cv=FALSE) {
    nd <- length(p)
    mat <- makematlasso(d,p,ndiv)
    y <- d$V2 - mean(d$V2)
    m <- glmnet(mat,y,alpha=alpha)
    ndim <- m$dim[2]
    pow <- matrix(0,nd,ndim)
    for (i in 1:ndim) {
        v <- m$beta[,i]
        for (j in 0:(ndiv-1)) {
            pow[,i] <- pow[,i] +
               v[(nd*j+1):(nd*(j+1))]^2
        }
    }
    nmin <- NULL
    gcv <- NULL
    if (cv) {
        gcv <- cv.glmnet(mat,d$V2,alpha=alpha)
        minl <- gcv$lambda.min
        nmin <- which.min(abs(
                m$lambda-gcv$lambda.min))
    }
    r <- list(pow=pow,p=p,lambda=m$lambda,
         nmin=nmin,m=m,gcv=gcv,mat=mat)
    class(r) <- c("lassopow",class(r))
    return(r)
}

plot.lassopow <- function(pow,n,...) {
    p <- pow$p
    pw <- pow$pow
    plot(p,pw[,n],typ="l",xlab="Period",
         ylab="Relative Power",...)
}

minper <- function(pow,n,num=1) {
    p <- pow$p
    pw <- pow$pow
    pp <- numeric()
    while (num > 0) {
        i <- which.max(pw[,n])
        pp <- c(pp,p[i])
        num <- num-1
        pw[i,n] <- 0
    }
    return(pp)
}

lassofit <- function(pow,x,n,ndiv=2) {
    d <- data.frame(V1=x,V2=numeric(length(x)))
    m <- pow$mat
    return(m %*% pow$m$beta[,n])
}

lassoexpect <- function(pow,x,n,ndiv=2) {
    d <- data.frame(V1=x,V2=numeric(length(x)))
    m <- makematlasso(d,pow$p,ndiv)
    return(m %*% pow$m$beta[,n])
}

# 2-D version

pergrmlasso <- function(d,fmin,fmax,div) {
    d$V1 <- d$V1 - mean(d$V1)
    d$V2 <- d$V2 - mean(d$V2)
    p <- seq(fmin,fmax,length=div+1)
    pl <- perlasso(d,1/p)
    las <- list(pow=pl$pow,p=pl$p,lambda=pl$lambda)
    r <- list(period=p,lassopow=las)
    class(r) <- c("PergrmLasso",class(r))
    return(r)
}

pergrm2dlasso2 <- function(d,fmin,fmax,div,len,step) {
    nd <- floor((diff(range(d$V1))-len)/step+1)
    lst <- list()
    st <- min(d$V1)
    cat("bins="); cat(nd); cat("\n")
    for (i in 1:nd) {
        t <- subset(d,V1 >= st+step*(i-1) & V1 < st+len+step*(i-1))
        if (nrow(t) > 10) {
            p <- pergrmlasso(t,fmin,fmax,div)
        } else {
            p <- NA
        }
        lst[[i]] <- p
    }
    return(lst)
}

getpergrmlasso2 <- function(d,start,end,fmin,fmax,div,len,step) {
# range should be expressed in frequency
    orgdata <- d
    d <- data.frame(V1=d$V1,V2=d$V2)
    f <- subset(d,V1 > start & V1 < end)
    r <- list(g=f,f=orgdata)
    g <- r$g
    f <- r$f
    dt <- 1
    lst <- pergrm2dlasso2(g,fmin,fmax,div,len,step)
    yy <- seq(fmin,fmax,length=div+1)
    nd <- floor((diff(range(g$V1))-len)/step+1)
    xx <- min(g$V1)+(0:(nd-1))*step+len/2
    r <- list(lst=lst,d=orgdata,f=f,g=g,xx=xx,yy=yy,len=len,dt=dt)
    class(r) <- c("Pergrm2lasso",class(r))
    return(r)
}

drawpgmsub <- function(pgm,pp,powmin,powmax,xlab,ylab) {
    d <- pgm$d
    xx <- pgm$xx
    yy <- pgm$yy
    dt <- pgm$dt
    len <- pgm$len
    freq <- pgm$freq

    def.par <- par(no.readonly = TRUE)

    mat <- matrix(c(1,1,2,2,2), 5, 1, byrow = TRUE)
    layout(mat)
    par(mar=c(2,5,0,4))
    par(oma=c(1,0,4,0))
    par(cex.lab=1.4)
    par(cex.axis=1.2)
    xlim <- c(min(xx)-dt*len/2,max(xx)+dt*len/2)

    plot(d$V1,d$V2,xlim=xlim,ylim=range(d$V2),typ="l",
         xaxs="i",pch=19,cex=0.3,xlab="",ylab="amplitude")

    par(mar=c(4,5,0,4))
    image(x=xx,y=yy,z=pp,xlim=xlim,
          zlim=c(powmin,powmax),col=gray.colors(100,start=1,end=0,gamma=2),
          xlab=xlab,ylab=ylab)

    par(def.par)
}                       

drawpgmlasso <- function(pgm,n,powmin,powmax,dup=0,
                         xlab="Time (d)",
                         ylab="Frequency (c/d)",
                         loglambda=TRUE) {
    if (!inherits(pgm,"Pergrm2lasso")) {
        stop("only Pergrm2lasso class can be drawn")
    }

    nn <- length(pgm$lst)
    pp <- matrix(0,nn,length(pgm$yy))
    zerovec <- rep(0,length(pgm$yy))
    if (dup == 0) {
        dwin <- 1
    } else {
        dwin <- 10^(-abs(((-dup):dup)/dup*10))
    }
    for (i in 1:nn) {
        if (is.na(pgm$lst[[i]][1])) {
            pw <- zerovec
        } else {
            if (loglambda == FALSE) {
                pw <- pgm$lst[[i]]$lassopow$pow[,n]
            } else {
                ni <- which.min(abs(log10(pgm$lst[[i]]$lassopow$lambda)-n))
                pw <- pgm$lst[[i]]$lassopow$pow[,ni]
            }
        }
        for (j in (-dup):dup) {
            if (i+j > 0 && i+j <= nn) {
                pp[i+j,] <- pp[i+j,] + dwin[j+dup+1]*pw
            }
        }
    }
    pp <- log10(pp)
    pp <- ifelse(pp > powmax,powmax,pp)
    drawpgmsub(pgm,pp,powmin,powmax,xlab,ylab)
}

\end{verbatim}

\section*{Acknowledgements}

This work was supported by JSPS KAKENHI Grant Number 21K03616.
This work was originally initiated under the grant
``Initiative for High-Dimensional Data-Driven Science through
Deepening of Sparse Modeling'' (2013--2018) from
the Ministry of Education, Culture, Sports, Science and
Technology (MEXT) of Japan, and I am grateful to
Toshiyuki Tanaka, Shiro Ikeda and Masato Okada for
comments on compressed sensing and for their encouragement
in application of compressed sensing to bioacoustics.
I am grateful to the observers who reported observations
of V1674 Her to VSNET.
I am grateful to the xeno-canto team and recordists
for providing the database and recordings available
to the public.

\section*{References}

  I provide two forms of the references section (for ADS
and as published) so that the references can be easily
incorporated into ADS.

\renewcommand\refname{\textbf{References (for ADS)}}

\newcommand{\noop}[1]{}\newcommand{\hyphalt}{-}

\xxinput{lasso2aph2.bbl}

\renewcommand\refname{\textbf{References (as published)}}

\xxinput{lasso2.bbl.vsolj}

\end{document}